\documentclass[debug,overfull]{epl}
\usepackage{amssymb}

\title{High-multipolar effects on the Casimir force: \\ the non-retarded limit}
\shorttitle{High-multipolar effects on the Casimir effect}
\author{Cecilia Noguez \and Carlos E. Rom\'an-Vel\'azquez \and Raul Esquivel-Sirvent \and Carlos Villarreal}
\shortauthor{Cecilia Noguez \etal}
\institute{ Instituto de F\'{\i}sica, Universidad Nacional Aut\'onoma
 de M\'exico, Apartado Postal 20-364, D.F. 01000,  M\'exico
 }
 

\begin{document}

\maketitle

\begin{abstract}
We calculate exactly the Casimir force or dispersive force, in the non-retarded limit, between a spherical nanoparticle and a substrate beyond the London's or dipolar approximation. We find that the force is a non-monotonic function of the distance between the sphere and the substrate, such that, it is enhanced by several orders of magnitude as the sphere approaches the substrate. Our results do not agree with previous predictions like the Proximity theorem approach.
\end{abstract}

%\section{Section title}
The technical and experimental advances associated to the investigation of micro and nano devices has boosted the research on forces acting at such scale, like Casimir and van der Waals (vW) dispersive forces. The origin of dispersive forces between atoms and macroscopic bodies may be attributed to electromagnetic interactions between their charge distributions induced by quantum vacuum fluctuations, even when they are electrically neutral. In a first approximation the charge distribution of neutral particles may be represented by electric dipoles. This dipole approximation was employed by London to calculate the non-retarded van der Waals interaction potential $V_{\rm vW} (z)$ between two identical polarizable molecules by using perturbation theory in quantum mechanics\cite{london}. He found that $V_{\rm vW}(z) \sim - \alpha^2/z^6$, where $\alpha$ is the polarizability of the particles and $z$ is the magnitude of the distance between them. 
In 1948, Casimir and Polder\cite{casypol} found a correction to the London-vW force by considering the influence of retardation, where at large distances the potential varies like $1/z^{7}$, instead of $1/z^{6}$. Furthermore, they found that the vW interaction could be attributed to the change of the zero-point energy\cite{casimir}, 
 \begin{equation}
  V_{\rm vW}(z) = {\cal E}(z) - {\cal E}(z \to \infty) = \frac{\hbar}{2}\sum_m[\omega_{m}(z) - \omega_{m}(z \to \infty)], \label{ener}
 \end{equation}
where $\omega_{m}(z)$ are the classical proper electromagnetic modes of the system. 

Later, Casimir studied a simpler problem\cite{casimir}: the change of the zero-point energy of two parallel conducting plates separated by a distance $z$, finding an interaction energy per unit area, ${\cal V}_{\rm C}(z) = -(\pi^2 \hbar c)/(720)(1/z^3), $ where $c$ is the speed of light. In 1956, Lifshitz\cite{lifshitz} extended the theory of Casimir to dielectric materials, and later on, it was shown that the Lifshitz's formula can be obtained from the zero-point energy of the interacting surface plasmons of the plates\cite{kampen,gerlach}. One of the first experimental evidences of the Casimir force was observed by Derjaguin and Abrikosova\cite{proximidad}. However, they measured the force between a glass plate and spherical lenses to avoid experimental difficulties inherent in keeping the plates parallel. To understand their measurements, Derjaguin and collaborators developed the so called Proximity Theorem to estimate the Casimir force per unit area between two spheres of radii $R_1$ and $R_2$, in terms of the energy per unit area between parallel plates, ${\cal V}_{\rm C}(z)$. Assuming that the Casimir force on a small area of one sphere is due to locally ``flat'' surfaces on the other sphere, it is found  that 
\[F_{\rm PT}(z) = 2 \pi \left(\frac{R_1R_2}{R_1 + R_2}\right) {\cal V}_{\rm C}(z).\]
This theorem holds when $z \ll R_1, R_2$, and it is not clear up to what limit this approach is valid. In the limit when  $R_1 =R$, and $R_2 \to \infty$, the problem reduces to the case of a sphere of radius $R$ and a flat plate, that yields to $F_{\rm PT}(z) = 2 \pi R {\cal V}(z)$. If the sphere and plate are perfect conductors, \[ F_{\rm PT}(z) = -\frac{\pi^3 \hbar c}{360} \frac{R}{z^3}, \] while the lack of retardation implies a behavior of $ \sim 1/z^{2}$. 

Casimir and Polder\cite{casypol} also studied the interaction of a neutral atom with a perfectly conducting plate, such that, the force behaves as $F \sim -\alpha/z^5$ at large distances and as $ -\alpha/z^4$ at small distances (without retardation effects). The same behavior can be also obtained calculating the change of the zero-point energy between a induced dipole moment on a sphere of polarizability $\alpha$ with its own image dipole in a plate. Contrary to what one may expect, this result differs from the one obtained using the Proximity Theorem by a factor of $1/z^2$. Therefore, an exact calculation of the dispersive force between a sphere and a planar surface becomes essential.  In this direction, Langbein\cite{langbein} used perturbation theory to calculate the dispersion forces between two dielectric spheres by considering that they are composed of fluctuating electric dipoles. He considered that the dipoles of sphere A interact among them, giving rise to a screened field that influenced the dipoles of sphere B, which also induced a reaction field on A, and so on. He expanded the field fluctuations in terms of spherical harmonics obtaining an infinite Taylor series in terms of $R/z$. Langbein only evaluated explicitly some upper and lower bounds for the interaction energy by considering: (i) few terms of the Taylor series, (ii) approximate expressions for the polarizabilties of the spheres, and (iii) constant dielectric functions for frequency ranges lower than a cutoff frequency. Under those approximations, the upper energy-bound is consistent with the Proximity Theorem. 

More recently, Johansonn and Apell\cite{appel} calculated the vW force between a sphere and a semi-infinite slab to estimate the errors involved in the Proximity Theorem. They calculated the electromagnetic stress tensor in terms of the electric field correlation associated to the response to an electric dipole by determining the Green's function of the Poisson's equation in bispherical coordinates. They concluded that for small separations  the behavior of the attractive force is consistent with the Proximity Theorem. However, one drawback in this formalism, relevant for the purposes of the present paper, is that in bispherical coordinates the section surfaces become planar (a point) for small (large) values of $z$, so that arbitrary values of ratio of $z/R$  cannot  be considered. Ford also investigated the Casimir force of a dielectric sphere and a wall, within the dipole approximation\cite{ford}. He found that the relative force could oscillate from attractive to negative depending on the distance of the sphere to the wall. However, this oscillatory behavior has not been observed experimentally\cite{lamoraux,mohideen,chan,decca} up to date. 

In this paper, we study the dispersive force between a sphere and a semi-infinite substrate in the non-retarded limit. We calculate the interaction energy as the difference of the zero-point energy when the sphere of radius $R$ is at a distance $z$ from the substrate, and when the sphere is at infinite. The proper electromagnetic modes of the sphere-plate system are calculated taking into account all the multipolar excitations, instead of dipolar ones. We find these modes using a spectral representation formalism\cite{ceci2}. The spectral representation formalism has the advantage that separates the contribution of the dielectric properties of the sphere from the contribution of its geometrical properties, such that, it is possible to perform a systematic study of the system. 

%We show that the dependance of the interaction energy with $z$ is not monotonic, such that, the force is enhanced several orders of magnitude as the sphere approaches the substrate.

Our model consist of a nanometric-size sphere of radius $R$ located at a minimum distance $z$ from a substrate, with local dielectric functions $\epsilon_{\rm sph}(\omega)$, and $\epsilon_{\rm sub} (\omega)$, respectively. In the non-retarded limit, $R$ and $z$ must be smaller than a characteristic length, $l$, of the system (for a metallic sphere $l=c/\omega_p$, where $\omega_p$ is the plasma frequency of the sphere). We assume, like in vW, that the fluctuations of vacuum induce a charge distribution on the sphere which also induces a charge distribution in the substrate, such that, the $lm$-th multipolar moment on the sphere is given by  \begin{equation}
Q_{lm}  = -\alpha_{lm} \frac{(2l+1)}{4 \pi}  \big\{ V_{lm}^{\rm vac} + V_{lm}^{\rm sub} \big\}, \nonumber 
\end{equation}
where $V_{lm}^{\rm vac}$ is the exciting field associated to the quantum vacuum fluctuations at the zero-point energy, $V_{lm}^{\rm sub}$ is the induced field due to the presence of the substrate, and $\alpha_{lm}$ are the polarizabilities of the sphere\cite{claro}. Here, the $lm$-th multipole moment induced in the sphere is defined as 
\[Q_{lm}=\int_{v_s} r'^l \rho({\bf r'}) Y_{lm}(\theta', \varphi')\,d^3r' , \] 
where $\rho ({\bf r'})$ is the charge density in the sphere, $Y_{lm}$ are the spherical harmonics, and the integral is performed over the sphere's volume $v_s$. The $lm$-th component of the field can be written using a multipolar expansion\cite{claro}, that yields to
\begin{equation}
Q_{lm} = \alpha_{lm} \frac{-(2l+1)}{4 \pi} \bigg\{ V_{lm}^{\rm vac} + \sum_{l', m'} (-1)^{m^{^{\prime }}+l^{^{\prime }}}  A_{lm}^{l'm'} {\hat{Q}}_{l'm'} \bigg\},  \label{q}
\end{equation}
where $\hat{Q}_{l'm'}$ is the $l'm'$-th induced multipolar moment in the substrate  which is located at ${\bf r}= (2(z+R),\theta=\pi,\varphi)$ from the center of the sphere, and $A_{lm}^{l'm'}$ is the matrix that couples the interaction between the multipolar distribution on the sphere and substrate\cite{claro}. The induced $l'm'$-th multipolar moment in the substrate or the ``image-multipole'' moment  is related with the $Q_{lm}$ on the sphere by 
\begin{equation}
\hat{Q}_{l'm'} = (-1)^{l'+m'} f_c(\omega) Q_{l'm'}, \label{qp}
\end{equation}
where $f_c(\omega)$ is a contrast function that together with $(-1)^{l'+m'}$ allows to satisfy the boundary conditions of the electric field on the plate. The contrast function is given by
\begin{equation}
f_c(\omega) = \frac{1 - \epsilon_{\rm sub}(\omega)}{1 + \epsilon_{\rm sub}(\omega)}, 
\end{equation}
and depends only on the dielectric properties of the plate. Substituting  eq.~(\ref{qp}) in eq.~(\ref{q}), one finds 
\begin{equation}
-\sum_{l'm'} \left[\frac{4 \pi}{(2l'+1)}\frac{ \delta_{ll'} \delta_{mm'} }{\alpha_{l'm'}} + f_c (\omega) A_{lm}^{l'm'} \right] {Q}_{l'm'} = V_{lm}^{\rm vac}.  \label{q2}
\end{equation}
The proper electromagnetic modes of the sphere-substrate system satisfy Eq.~(\ref{q2}), and are independent of the exciting field. Then, the normal frequencies can be obtained when the determinant of the matrix in the left-side of Eq.~(\ref{q2}) is equal to zero. One can solve Eq.~(\ref{q2}) by a self-consistent procedure, or inverting $L(2L+1)$ complex matrices of $L\times L$ dimension, where $L$ is the largest order of the multipolar expansion. To avoid these cumbersome procedures, we use instead a spectral representation method to find the proper electromagnetic modes. The discussion of the above method can be found in Ref.\ref{ceci2}.

Suppose that the sphere is homogeneous, such that, its polarizabilities are independent of the index $m$, and are given by\cite{claro} 
\begin{equation}
\alpha_l (\omega) = \frac{l[\epsilon_{\rm sph}(\omega) - 1]} {l [ \epsilon_{\rm sph}(\omega) + 1] + 1} R^{2l+1}. \label{alfa}
\end{equation}
Let us rewrite such polarizabilities like
\begin{equation}
\alpha_l (\omega) =  \frac{n_0^l}{n_0^l - u(\omega)} R^{2l+1}, \label{alfa2}
\end{equation}
where $u(\omega) = [1 - \epsilon_{\rm sph}(\omega)]^{-1}$ and $n_0^l = l/(2l+1)$. The poles of  Eq.~(\ref{alfa2}), $u(\omega^l_0) = n_0^l$, yield the frequencies of the proper electromagnetic modes of the isolated sphere. Therfore, we define $u(\omega)$ as the spectral variable. Note that in Eq.~(\ref{alfa2}), we have separated the material from the geometrical properties of the sphere. 

Now, using  $u(\omega)$, we rewrite Eq.~(\ref{q2}) as \begin{equation} 
\sum_{\mu'} \left[ - u(\omega) \delta_{\mu \mu'}  +  {H}^{\mu '}_{\mu}  \right] {x}_{\mu'} = b_{\mu}, \label{h2}
\end{equation}
where we have simplified the notation by writing 
\[ \mu \equiv (l,m), \quad   x_{\mu} = \frac{Q_{lm}} { (l R^{2l+1})^{1/2}}, \quad {\rm and} \quad b_{\mu} = - \frac{(l R^{2l+1})^{1/2}} {4 \pi} V_{lm}^{\rm vac}, \]
and
\begin{equation}
H_{\mu}^{\mu'} = n_{l'}^0  \delta_{\mu \mu'} + f_c \frac{(R^{l+l'+1})} {4 \pi} (ll')^{1/2} A_{\mu}^{\mu'}. \label{h}
\end{equation}
It was shown in Ref.\ref{clar} that, $A_{\mu}^{\mu'}$ is a symmetric matrix that depends on the distance between the centers of the``image-multipole'' and sphere as, $1/[2(z+R)]^{l+l'+1}$, such that, ${\mathbb H}$ is dimensionless and symmetric, and depends only on the geometry of the system. Consider the case when the contrast factor function $f_c(\omega)$ is real (we can perform a rotation to the imaginary axis  ($\omega \to i\omega$) to get a real dielectric function), then ${\mathbb H}$ is also real. We can always find a unitary matrix ${\mathbb U}$ that diagonalizes ${\mathbb H}$, such that,
\[ \sum_{\mu \mu'}  (U_{\nu}^{\mu})^{-1}H_{\mu}^{\mu'}U_{\mu'}^{\nu'}  =  4 \pi n_{\nu} \delta_{\nu \nu'},\] 
being $n_{\nu}$ its eigenvalues. The solution of Eq.~(\ref{h2}) is given by $ x_{\mu} = - \sum_{\mu'} G_{\mu}^{\mu'} b_{\mu'}$, 
where $ G_{\mu}^{\mu'}$ is a Green's operator, whose $\mu \mu'$ element can be written in terms of $U_{\nu}^{\mu}$ and $n_{\nu}$, as
\begin{equation}
G^{\mu}_{\mu'}(u) = \sum_{\mu''}  \frac{U_{\mu''}^{\mu}(U_{\mu'}^{\mu''})^{-1}}{u-n_{\mu''}}. \label{green}
\end{equation}
Now, the poles of $G^{\mu}_{\mu'}(u)$ yield the proper frequencies of the system. Finally, the interaction energy is calculated according to Eq.~(\ref{ener}).

The results presented here are calculated as follows. First, we construct and diagonalize the matrix $\mathbb H$ for a given $z/R$, choosing a maximum value $L$ of multipolar excitations that ensures convergence in the interaction energy. Then,  we consider an explicit expression for the dielectric function of the sphere, and calculate the proper electromagnetic modes trough the relation $u(\omega^{l}_{m}) = n_{m}^{l}$. Once we have $\omega^l_m$, we calculate the energy according with Eq.~(\ref{ener}). The largest order of the multipolar expansion considered in this work was $L=2000$ for $z/R = 0.001$. As a case study, we use the Drude model for the dielectric function of the sphere, therefore, $u(\omega) =  [\omega(\omega + i/\tau)]/\omega_p^2$, where $\omega_p$ is the plasma frequency and $\tau$ is the relaxation time. We present results for aluminum (Al) spheres  with $\hbar \omega_p = 15.80$~eV, and $(\tau \omega_p)^{-1} = 0.04$. The substrate is sapphire whose dielectric function is real and constant in a wide range of the electromagnetic spectrum\cite{sapphire}. 

\begin{figure}[tbh] 
\onefigure{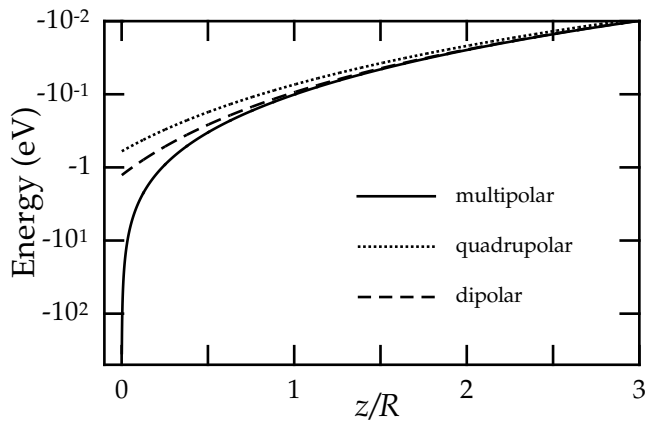}
\caption{Energy as a function of the ratio $z/R$, when all multipolar effects are taken into account (solid line), as well as results when only  quadrupolar (dashed line), and dipolar (doted line) effects are considered.}
\label{figura1}
 \end{figure}
 
In fig.\ref{figura1}, we observe that the energy shows a power law of $(z/R)^{-3}$ when only dipolar effects are considered while the curve corresponding to quadrupolar excitations shows three different regions: (i) when $z/R > 7$ the energy shows a power law of $(z/R)^{-3}$, that means that only dipole-dipole interactions are important, (ii) when $ 2 < z/R < 7$ the energy shows a power law of $(z/R)^{-4}$, indicating that dipole-quadrupole interactions are dominant, and (iii) when $z/R < 2$ the energy shows a power law of $(z/R)^{-5}$, such that quadrupole-quadrupole interactions become important. At small distances, the energy of the system can increase up to three orders of magnitude when all the contributions from high-multipolar moments are considered, as compared with the energy when only dipolar or quadrupolar moments are taken into account. Then, the exact energy curve shows a power law of $(z/R)^{-3}$ when $z/R > 7$, when $ 2 < z/R < 7$ the energy shows the same behavior as the curve where quadrupolar effects are included, and at smaller distances ($z/R < 1$) the energy increases sharply as $z$ decreases, and it is not possible to assign a power law. This means that the interaction energy is not a monotonic function of $z$ like in the Casimir and Polder, and Proximity Theorem approaches. 

\begin{figure}[tbh] 
\onefigure{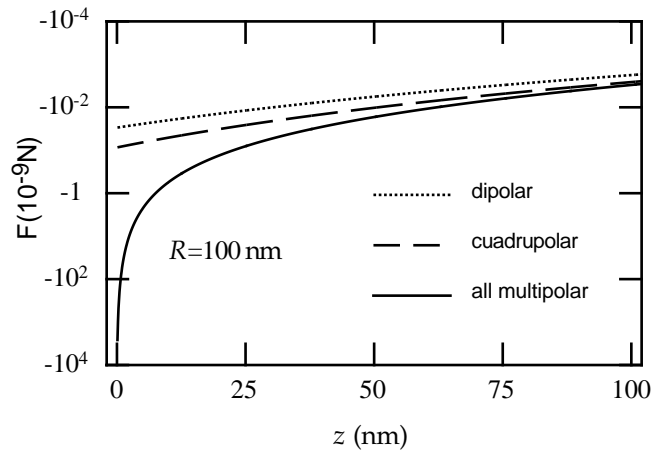}
\caption{Casimir force as a function of the minimum distance $z$ for a sphere with $R=50$~nm.}
\label{figura2}
 \end{figure}

In fig.\ref{figura2}, we show the Casimir force for an Al sphere over sapphire calculated as $F = - d{\cal E}/dz. $ We show the force when all multipolar interactions are taken into account, as well as up to dipolar, and up to quadrupolar interactions are considered. In agreement with the results for the energy, we observed for the force that multipolar effects become evident when the minimum distance between the sphere and the substrate is smaller than $R$. As the sphere approaches the substrate, the attractive force suddenly increases up to four orders of magnitude as compare with the dipolar interactions. At large distances ($z > 2R$) the force can be obtained exactly if only up to quadrupolar interactions are considered. We also obtained that for $z > 7R$ the interaction between the sphere and the substrate can be modeled using only the dipolar approximation\cite{ceci2}. 
 
Shih and Parsegian\cite{shih} measured the deflection of atomic beams by gold surfaces to study the Casimir-Polder interaction. However, they  obtained results that are inconsistent with Casimir and Polder, and also with Proximity Theorem, observing larger deflections. Also, recent experiments using the sphere-plate configuration have been performed to measure the Casimir force\cite{lamoraux,mohideen,chan,decca} and their interpretation has relied in the Proximity Theorem. However, to compare experimental data to theory, it has been needed to make two significant modifications to the Casimir force within the proximity theorem approximation. The first one was to employ the Lifshitz formula\cite{lifshitz} to calculate the energy density of parallel plants, ${\cal V}_{\rm C}(z)$. Then, they employ the Proximity Theorem to calculate the force, $F_{\rm PT}(z)$. The second modification uses the fact that they measured a larger attractive force than the one predicted by the Proximity Theorem. They attributed the deviations to the roughness of the surface which tend to increase the attraction force like, $F(z) = F_{\rm PT}(z) \left[ 1 + 6(A_r/z)^2 + 15 (A_r/z)^4 + \cdots \right] $, where the constant $A_r$ depends on the model of the surface roughness\cite{maradudin,bezerra}. In summary, the measurements of Shih and Parsegian\cite{shih}, as well as other experiments\cite{chan}, indicate that dispersive forces between a polarizable atom or spherical particle and a planar substrate involves more complicated interactions than the simple dipole model of Casimir and Polder or the Proximity Theorem approximation.  

In conclusion, we have shown that the dispersive force between a sphere and a substrate is a non-monotonic function of the distance between bodies, showing that the force is enhanced by several orders of magnitude as a nanoparticle approaches the plate.  On the other hand, at large separations the dipolar term dominates the interaction energy, like in the Casimir and Polder model. The increment of the force at small separations could explain the physical origin of the large deviations observed in the deflection of atomic beams by metallic surfaces, as well as some instabilities detected in micro and nano devices. However, specific experiments have to be perform to prove the latter. Our results are in contradiction with the Proximity Theorem. As we pointed out before, the Proximity Theorem does not take into account the complete geometry of the system. This is more clear if one calculates the zero-point energy of the surface modes within the dipolar approximation. In the non-retarded limit, the longitudinal surface mode of a dielectric plate is $\omega_{\rm plate} = \omega_p/\sqrt{2}$, while for a sphere the surface mode is $\omega_{\rm sphere} = \omega_p/\sqrt{3}$. It is evident that it is not possible to obtain the sphere's surface mode from a sum of plate's surface modes, or vice-versa. Then, the proximity theorem is not able to include all the effects inherent to the sphere-plate geometry. 

\acknowledgments
We would like to thank to Prof. R.G. Barrera for useful discussions. This work has been partly financed by CONACyT grant No.~36651-E and by DGAPA-UNAM grants No.~IN104201 and IN107500.

\end{document}